\documentclass{emulateapj}

\usepackage{natbib,amsmath}

\newcommand{\cb}{MS~1512-cB58}
\newcommand{\aval}{-0.92 \pm 0.05}
\newcommand{\aonly}{-0.92}
\newcommand{\bval}{1.41 \pm 0.10}
\newcommand{\bonly}{1.41}
\newcommand{\mwtrend}{[M/H]~$= \aonly + \bonly \log(W_{1526} / 1 {\rm \AA})$}
\newcommand{\msol}{M_\odot}

\newcommand{\kms}{km~s$^{-1}$}
\newcommand{\cm}[1]{\, {\rm cm^{#1}}}
\newcommand{\mkms}{{\rm \; km\;s^{-1}}}
\newcommand{\delv}{$\Delta v_{90}$}
\newcommand{\mdelv}{\Delta v_{90}}

\newcommand{\wsi}{$W_{\rm 1526}$}
\newcommand{\mwsi}{W_{\rm 1526}}
\newcommand{\wciv}{$W_{\rm CIV}$}

\newcommand{\lya}{Ly$\alpha$}

\newcommand{\N}[1]{{N({\rm #1})}}
\newcommand{\sci}[1]{{\rm \; \times \; 10^{#1}}}
\newcommand{\mnhi}{N_{\rm HI}}

\newcommand{\nhi}{$N_{\rm HI}$}

\def\fnhi{$f(\mnhi)$}

\begin{document}

\title{On the Nature of Velocity Fields in 
High $z$ Galaxies}

\author{Jason X. Prochaska\altaffilmark{1},
        Hsiao-Wen Chen\altaffilmark{2}, 
	Arthur M. Wolfe\altaffilmark{3},
	Miroslava Dessauges-Zavadsky\altaffilmark{4},
        Joshua S. Bloom\altaffilmark{5}
}
\altaffiltext{1}{Department of Astronomy and Astrophysics, 
UCO/Lick Observatory;
University of California, 1156 High Street, 
Santa Cruz, CA 95064; xavier@ucolick.org}
\altaffiltext{2}{Department of Astronomy; University of Chicago;
5640 S. Ellis Ave., Chicago, IL 60637; hchen@oddjob.uchicago.edu}
\altaffiltext{3}{Department of Physics, and Center for Astrophysics and 
Space Sciences, University of California, San Diego, C--0424, La Jolla, 
CA 92093-0424}
\altaffiltext{4}{Observatoire de Gen\`eve, 51 Ch. des Maillettes, 
1290 Sauverny, Switzerland}
\altaffiltext{5}{Department of Astronomy, 601 Campbell Hall, 
University of California, Berkeley, CA 94720-3411}

\begin{abstract}
We analyze the gas kinematics of 
damped \lya\ systems (DLAs) hosting high $z$ gamma-ray
bursts (GRBs) and those toward quasars (QSO-DLAs) focusing on three
statistics:  
(1) \delv, the velocity interval encompassing 90$\%$ of the total
optical depth, 
(2,3) \wsi\ and \wciv, the rest equivalent widths of 
the \ion{Si}{2}~1526 and \ion{C}{4}~1548 transitions.
The \delv\ distributions of the GRB-DLAs and QSO-DLAs 
are similar, each has median $\mdelv \approx 80 \mkms$ 
and a significant tail to several hundred \kms.
This suggests comparable galaxy masses
for the parent populations of GRB-DLAs and QSO-DLAs and we infer
the average dark matter halo mass of GRB galaxies 
is $\lesssim 10^{12} M_\odot$.   
The unique configuration of GRB-DLA sightlines and the presence
(and absence) of fine-structure absorption together give 
special insight into the nature of high $z$, protogalactic velocity fields.
The data support a scenario
where the \delv\ statistic reflects dynamics in the interstellar medium
(ISM) and \wsi\ traces motions outside the ISM (e.g.\ halo gas, galactic-scale
winds). 
The \wsi\ statistic and gas metallicity
[M/H] are tightly correlated, especially for 
the QSO-DLAs: [M/H]~=~$a + b \log(W_{1526}/1{\rm\AA})$
with $a = \aval$ and $b = \bval$.
We argue that the \wsi\ statistic primarily tracks dynamical motions in 
the halos of high $z$ galaxies and interpret  
this correlation as a mass-metallicity relation 
with very similar slope to the trend observed in local, low-metallicity
galaxies.
Finally, the GRB-DLAs exhibit systematically larger \wsi\
values ($>0.5$\AA) than the QSO-DLAs ($<\mwsi> \approx 0.5$\AA)
which may suggest galactic-scale outflows contribute to the largest
observed velocity fields.
%(iv) the GRB-DLA also show higher \wsi\ at higher metallicity.
%This would explain the larger
%\wsi\ values for GRB (which trace active star formation) and 
%also the observed correlation between \wsi\ and [M/H].
\end{abstract}

\keywords{quasars : absorption lines }

\section{Introduction}

Absorption-line spectra of quasars have revealed
thousands of high redshift 
galaxies to date \citep[e.g.][]{wolfe86,ss92,phw05,ppb06}.
At $z>2$, these galaxies are termed the damped \lya\ systems (DLAs),
absorbers with \ion{H}{1} column density $\mnhi \geq 2 \sci{20} \cm{-2}$.
Although a direct association between star forming galaxies 
and DLAs is difficult
to establish because the background quasar blinds our view \citep{mwf+02},
the large \nhi\ values of DLAs indicate large dark matter overdensities
(i.e.\ a virialized system).  Furthermore, all of the DLAs show
substantial
enrichment from heavy metals \citep{pettini94,pgw+03} and roughly half
show indications of ongoing star formation through the presence of 
\ion{C}{2}$^*$ absorption \citep{wpg03}.
Finally, stars form from neutral gas and the DLAs dominate
the atomic hydrogen reservoir at all redshifts \citep{wlf+95,phw05,rtn06}.

High resolution spectroscopy of DLAs yield precise measurements
of the column densities of resonance-line transitions in the ISM
of high $z$ galaxies \citep{wgp05}.
Aside from gas-phase abundances, however, it is difficult to
derive physical quantities from these observations  
(e.g.\ temperature, density).
The only other physical
characteristic easily studied is the gas kinematics.
High-resolution spectra resolve the line profiles of these
transitions into `clouds' which trace the velocity fields
in young, high $z$ galaxies \citep{pw97,pw98}.
Although it is difficult to reveal the nature
of these velocity fields (e.g.\ rotation, outflows, turbulence)
with individual 1-dimensional sightlines,
the distributions of kinematic characteristics 
provide powerful tests for scenarios of galaxy formation
\citep[e.g.][]{jp98,hsr98}.

\cite{pw97} performed a survey of the kinematic characteristics
of neutral gas
in DLAs and compared their observations against a variety of
simple scenarios.  These included rotating disks, clouds with random
(virialized) motions, and galactic infall.  Observed asymmetries
in the observed line-profiles favor rotational dynamics, yet the
observed distribution of velocity widths ($\Delta v$) has a median
too large to be accommodated within standard cold dark matter
cosmology.  As such, 
\cite{pw97} presented these observations as a direct challenge
to the basic picture of galaxy formation in a hierarchical universe
\citep[e.g.][]{kff96,mmw98}.
In response, theorists introduced additional velocity fields to
explain the observations \citep{hsr98,mcd99,nbf98,mps+01}.
Of particular interest was a model of merging protogalactic
clumps where the observed kinematics include
contributions from rotation, infall, and random motions \citep{hsr98}.
While the initial work established this model as a 
potential solution, recent evaluations using modern numerical
simulations and more accurate treatments of radiative transfer
have not confirmed its viability \citep{pw01,rnp+06}.
The research on kinematic characteristics in DLAs toward
quasars (hereafter QSO-DLAs),  therefore, tests
theories of galaxy formation and gas dynamics in the early universe.

Motivated by these results, we have investigated the
dynamics of gas probed
by the sightlines to the afterglows of gamma-ray bursts (GRBs).
Similar to quasars, the afterglows of GRBs provide bright, albeit
transient point-source beacons to the outer universe.
With follow-up spectroscopy, one can acquire data
at comparable signal-to-noise (S/N) and spectral resolution as
quasars \citep{vel+04,fdl+05,cpb+05,pcb+07}.
In turn, one can study the velocity fields of the gas 
close to the observed beacon. 
The sightline presumably intersects gas in the star forming
region encompassing the GRB, the neutral ISM surrounding it,
and gas external to the neutral ISM.

In this paper, we examine the 
kinematics of a modest sample of damped \lya\ systems associated
with the ISM surrounding gamma-ray bursts (GRB-DLAs).  
A principal goal is to characterize the velocity fields of gas
observed along these unique sightlines.
The data trace velocity fields generated by galactic
rotation, gravitational accretion, and any galactic or stellar
feedback processes.  
We will find that these observations reveal new insight for
interpretations of the kinematics of gas observed in quasar
absorption line studies.  Furthermore, we will present a comparison
of the GRB-DLA results with those from QSO-DLAs.
Comparisons between GRB-DLAs and QSO-DLAs
assess the relative galaxy masses
and the nature of velocity fields in young, star-forming galaxies.

This paper is organized as follows.
Section 2 presents the observational samples and describes the
experiment.  In $\S$~3, we define statistics for assessing
the gas kinematics.  $\S$~4 presents the principal results
of our analysis and we discuss and interpret these
results in $\S$~5.

\begin{figure}
\centerline{\epsfxsize=\hsize{\epsfbox{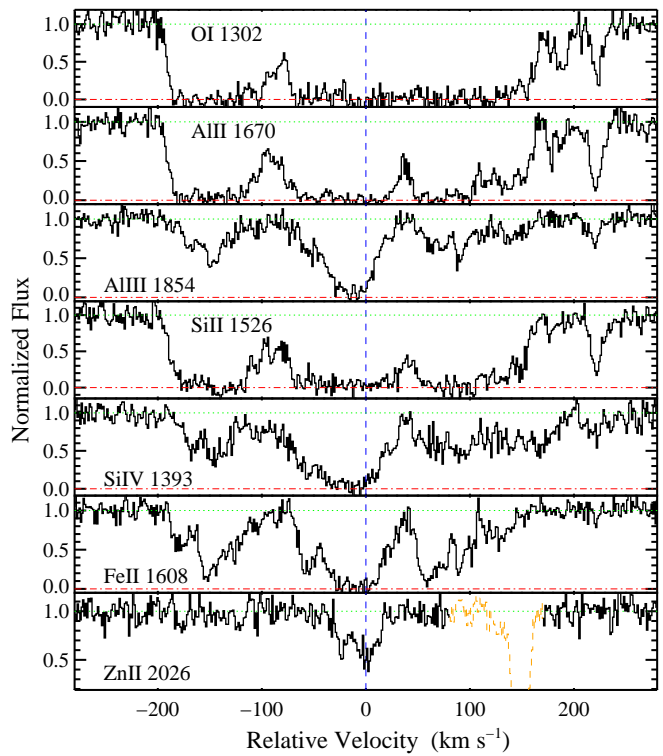}}}
\caption{Resonance-line profiles for the GRB-DLA
associated with GRB~050820 \citep{pcb+07}.  The velocity
$v=0$ corresponds to $z=2.61469$.  Note that the optical
depth in low-ion gas is dominated by a few clouds with velocity interval
$\mdelv \approx 50 \mkms$ (\ion{Zn}{2}~2026).
At the same time, there is absorption traced by strong
low-ion transitions (e.g.\ \ion{Si}{2}~1526)
to velocities of $\approx \pm 200 \mkms$.
}
\label{fig:exmpl}
\end{figure}

\section{Observational Data}

We have defined a sample of 
GRB-DLAs for kinematic analysis based on two primary criteria:
(i) the presence of a damped \lya\ system 
($\mnhi \ge 2\sci{20} \cm{-2}$)
or a low-ion column density that requires
\nhi\ exceed $2 \sci{20} \cm{-2}$ assuming solar metallicity;
(ii) spectra with sufficient signal-to-noise and resolution 
to study the gas kinematics.
The latter criterion is not especially strict because 
we will examine the gas
kinematics from equivalent width measurements which do not
require high spectral resolution. 
The sample is also limited to GRB-DLAs where
we could access the data or where precise equivalent
width measurements are reported in the literature.
Figure~\ref{fig:exmpl} shows an example set of line-profiles
for the GRB-DLA associated with GRB~050820 \citep{pcb+07}.  
These data were
acquired with the HIRES spectrometer on the Keck telescope
and have high spectral resolution $(R\approx 40,000)$.  
The weaker transitions such as \ion{Zn}{2}~2026 indicate that
the majority (90\%) of the gas is localized to one or two
`clouds' within a velocity interval of $\approx 50 \mkms$.  
In contrast, the stronger transitions are typically saturated 
and show absorption spanning 
nearly 400\,\kms.  In the following section, we introduce two
kinematic statistics to characterize these properties.
Table~\ref{tab:grbobs} lists the GRB-DLA comprising
our sample, describes the spectral observations, and
lists the \nhi\ values and metallicities as described
in \cite{pcd+07}.

\begin{deluxetable*}{lcccccccccc}
\tablewidth{0pc}
\tablecaption{GRB-DLA SAMPLE\label{tab:grbobs}}
\tabletypesize{\tiny}
\tablehead{\colhead{GRB} &\colhead{RA} & \colhead{DEC} & \colhead{$z_{GRB}$} &
\colhead{log \nhi} & 
\colhead{f$_{\rm mtl}^a$} & \colhead{[M/H]} &
\colhead{$\sigma({\rm [M/H]})$} &
\colhead{Instrument} & 
\colhead{$R$} & \colhead{Ref}}
\startdata
GRB990123&15:25:30.34&+44:45:59.1&1.600&$22^{b}$& 12&$-0.97$&&Keck/LRIS&1,000&1\\
GRB000926&17:04:09.00&+51:47:10.0&2.038&$21.30^{+0.25}_{-0.25}$&  2&$-0.17$&0.29&Keck/ESI&5,000&2\\
GRB010222&14:52:12.55&+43:01:06.2&1.477&$22^{b}$& 12&$-1.30$&&Keck/ESI&5,000&3\\
GRB011211&11:15:17.98&-21:56:56.2&2.142&$20.40^{+0.20}_{-0.20}$& 11&$-1.36$&&VLT/FORS2&1,000&1,4\\
GRB020813&19:46:41.87&-19:36:04.8&1.255&$22^{b}$& 12&$-1.17$&&Keck/LRIS&1,000&5\\
GRB030226&11:33:04.93&+25:53:55.3&1.987&$20.50^{+0.30}_{-0.30}$& 11&$-1.31$&&Keck/ESI&5,000&6\\
GRB030323&11:06:09.40&-21:46:13.2&3.372&$21.90^{+0.07}_{-0.07}$& 12&$-0.87$&&VLT/FORS2&1,000&7\\
GRB050401&16:31:28.82&+02:11:14.8&2.899&$22.60^{+0.30}_{-0.30}$& 12&$-1.57$&&VLT/FORS2&1,000&8\\
GRB050505&09:27:03.20&+30:16:21.5&4.275&$22.05^{+0.10}_{-0.10}$& 11&$-1.25$&&Keck/LRIS&1,000&9\\
GRB050730&14:08:17.14&-03:46:17.8&3.969&$22.15^{+0.10}_{-0.10}$&  4&$-2.26$&0.14&Magellan/MIKE&30,000&10\\
GRB050820&22:29:38.11&+19:33:37.1&2.615&$21.00^{+0.10}_{-0.10}$&  4&$-0.63$&0.11&Keck/HIRES&30,000&11\\
GRB050904&00:54:50.79&+14:05:09.4&6.296&$21.30^{+0.20}_{-0.20}$& 11&$-1.10$&&Subaru/FOCAS&1,000&12\\
GRB050922C&19:55:54.48&-08:45:27.5&2.199&$21.60^{+0.10}_{-0.10}$&  4&$-2.03$&0.14&VLT/UVES&30,000&13\\
GRB051111&00:08:17.14&-00:46:17.8&1.549&$22^{b}$& 12&$-0.96$&&Keck/HIRES&30,000&14,11\\
GRB060206&13:31:43.42&+35:03:03.6&4.048&$20.85^{+0.10}_{-0.10}$&  4&$-0.85$&0.18&WHT/ISIS&4,000&15\\
GRB060418&15:45:42.40&-03:38:22.80&1.490&$22^{b}$&  2&$-1.65$&1.00&Magellan/MIKE&30,000&11\\
\enddata
\tablenotetext{a}{Flag describing the metallicity measurement [M/H]: 0=No measurement; 1=Si measurement; 2=Zn measurement;
3=Combination of limits; 4=S measurement; 11=Lower limit from [$\alpha$/H]; 12=Lower limit from Zn}
\tablenotetext{b}{Because $z_{GRB} \le 1.6$, \lya\ was not observed.  We have set $\mnhi = 10^{22} \cm{-2}$, the median value of GRB-DLA.}
\tablerefs{
1: \cite{sff03};
2: \cite{cgh+03};
3: \cite{mhk+02};
4: \cite{vsf+06};
5: \cite{bsc+03};
6: \cite{sbp+06};
7: \cite{vel+04};
8: \cite{wfl+06};
9: \cite{bpck+05};
10: \cite{cpb+05};
11: \cite{pcb+07};
12: \cite{kka+06};
13: \cite{pwf+07};
14: \cite{pcb06};
15: \cite{fsl+06}}
 
\end{deluxetable*}

Our analysis will frequently draw comparisons between 
GRB-DLA measurements and QSO-DLA values.  
For the latter, we will consider
two samples of QSO-DLAs, one a subset of the other.
The complete sample is drawn from echelle and echellette
observations of quasars acquired with the HIRES \citep{vogt94}
and ESI \citep{sheinis00} spectrometers at the Keck Observatory
and the UVES \citep{uves} spectrometer at the VLT Observatory.
The data are summarized in these papers: \cite{shf06,ledoux06,dz06,pwh+07}.
We will restrict the samples to $z_{DLA} > 1.6$ and DLAs which are
greater than $3000\mkms$ from their background quasar.  
The full dataset is a heterogeneous
sample of QSO-DLAs including systems selected on
the basis of strong metal-lines \citep{shf06} or targeted 
for H$_2$ absorption \citep{ledoux03}.  As such, \cite{pwh+07} have
noted that the 
\ion{H}{1} distribution \fnhi\ of this full sample
does not follow the statistical distribution\footnote{The
QSO-DLAs observed at high spectral resolution have systematically
fewer systems with $\mnhi \approx 2\sci{20} \cm{-2}$, presumably
because the observers wished to avoid including absorbers
right at the threshold.}
derived from a random survey of background quasars \citep[e.g.][]{phw05}.
Therefore, we define a pseudo-statistical sample of QSO-DLAs
by restricting the list to the QSO-DLAs compiled by \cite{pgw+03}.  
Although the DLA in this subset also do not follow the \fnhi\ 
distribution of a random sample, they were selected only on the basis of 
a large \ion{H}{1} column density, i.e.\ independent of any
kinematic characteristic or chemical abundance.

We have analyzed the kinematic properties from our own HIRES and 
ESI observations and supplemented these measurements with the results
presented in \cite{ledoux06} from UVES data acquired with the VLT.  
For equivalent width measurements,
however, we have derived values only from our HIRES and ESI
observations \citep{pwh+07,shf06}.  These values have typical
statistical uncertainties of less than 0.02\AA.

\begin{figure}
\centerline{\epsfxsize=\hsize{\epsfbox{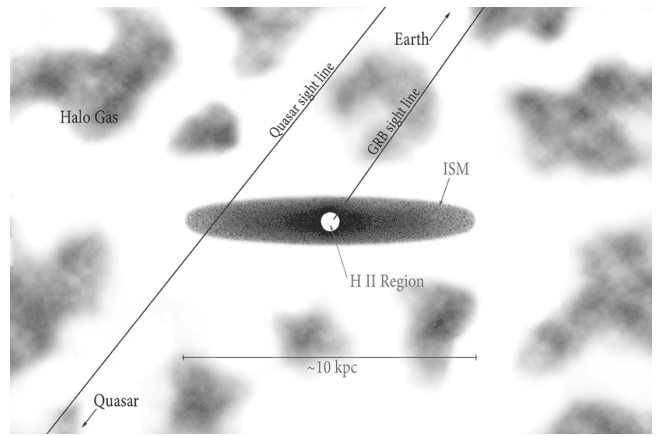}}}
\caption{This cartoon illustrates the likely differences
between QSO-DLA and GRB-DLA sightlines.  The former have randomly
intersected a foreground galaxy. These QSO-DLA sightlines correspond
to a cross-section selected sample and should preferentially
intersect the outer regions of the ISM in high $z$ galaxies.
In contrast, the GRB-DLA sightlines are constrained to originate from within
the ISM of their host galaxies, presumably the \ion{H}{2} region
produced by massive stars in a star-forming region.  These
GRB-DLA sightlines are expected (and observed) to originate within
the inner few kpc of the ISM.
}
\label{fig:cartoon}
\end{figure}

In a previous paper \citep{pcd+07}, we 
discussed why the parent populations of galaxies
hosting QSO-DLAs and GRB-DLAs may be different (e.g.\ gas cross-section
selected versus current SFR selected) and how the sightline
configurations are fundamentally distinct.
The GRB-DLAs do exhibit larger \nhi, metallicity, $\alpha$/Fe, and
depletion levels than QSO-DLAs \citep{pcd+07}, but we argued that the differences
do not require distinct parent populations of galaxies.
Instead, the observations imply sightlines with smaller average
impact parameter through high $z$ galaxies.
This conclusion is expected for a sample selected by gas
cross-section (QSO-DLA) versus ones restricted to originate in 
star-forming regions (GRB-DLA).
The cartoon in Figure~\ref{fig:cartoon} illustrates these ideas.
While QSO-DLAs preferentially penetrate the outer regions of the
ISM,  the GRB-DLA sightlines originate
within their host galaxies, presumably within an \ion{H}{2}
region generated by its progenitor and other O and B stars.
This has two key implications for the gas kinematics: 
(i) the GRB-DLA observations will probe only a fraction of the 
velocity field along the full sightline;
(ii) each GRB-DLA sightline is biased to intersect its own star
forming region.  We expect these regions to have very small
cross-section to QSO-DLA sightlines \citep{zp06} and to be 
rarely probed by such samples.
In this respect, GRB-DLAs may open a new window into the velocity
fields of high $z$ galaxies \citep[c.f.][]{pss+01}.

\section{Kinematic Diagnostics}
\label{sec:stats}

With high resolution observations, one resolves the
absorption-line profiles of unsaturated transitions
into individual components, frequently termed `clouds'.
The clouds in QSO-DLAs
have typical Doppler widths of $b = \sqrt{2} \sigma \approx 5 - 10 \mkms$ 
\citep[e.g.][]{dz06}.  The dispersion results from macroscopic motions
(e.g.\ turbulence) not thermal broadening; the latter
imply temperatures that would collisionally ionize the gas.
The GRB-DLAs show clouds with similar characteristics that
comprise the observed line-profiles 
\citep[Figure~\ref{fig:exmpl};][]{fdl+05,cpb+05,pcb+07}.

One can characterize the velocity field along the sightline
by tracing the motions of these clouds.  
Presently, we do not have measurements
of the systemic redshifts for our sample of GRB host galaxies (e.g.\ via
nebular lines). 
Therefore, we will primarily discuss relative velocities. 
Standard practice is to define a velocity width \delv\ as the interval
which encompasses 90$\%$ of the total optical depth of the
gas.  This definition was introduced in part
to limit a single, weak cloud or statistical fluctuations from 
dominating the statistic \citep{pw97}.  
Following the approach for QSO-DLAs \citep{pw97,pw98,pw01,ledoux06}, 
we measure the \delv\ statistic from
a single, unsaturated low-ion transition\footnote{Note that \delv\ is 
insensitive to the specific transition analyzed provided
the line is unsaturated and is associated with a low-ion
\citep{pw97}.} which, ideally, has
relatively high S/N.  For the QSO-DLA we demand
S/N~$> 15 {\rm pix^{-1}}$ but have relaxed this requirement
for the GRB-DLA because of the generally poorer data quality. 
To resolve the optical-depth profile, however, it is necessary
to restrict the sample to echelle or echellette observations.
As an example of the procedure, consider the line-profiles
in Figure~\ref{fig:exmpl} where we 
measure $\mdelv = 55 \mkms$ using the \ion{Zn}{2}~$\lambda 2026$ transition.  
Table~\ref{tab:kin} summarizes the measurements
for the full GRB-DLA sample.
%One can measure the same quantity for high-ion gas
%\hdelv\ using a \ion{Si}{4} or \ion{C}{4} transition \citep{wp00a}.
%These transitions, however, are frequently saturated (especially in 
%GRB-DLA) and \hdelv\ may not correspond to precisely 90$\%$
%of the high-ion gas.  Nevertheless, this quantity traces the 
%dynamics of highly ionized gas in the DLA and may be even  
%more sensitive to the gravitational potential and/or 
%feedback processes \citep{wp00b,mps+03}.

\begin{deluxetable}{lccccc}
\tablewidth{0pc}
\tablecaption{GRB-DLA KINEMATIC CHARACTERISTICS SUMMARY\label{tab:kin}}
\tablehead{\colhead{GRB} &
\colhead{$\lambda_{\rm low}$} & \colhead{$\Delta v_{\rm low}$} &
\colhead{W$_{1526}$} & \colhead{$W_{\rm CIV}$}\\
& (\AA) & (\kms) & (\AA) & (\AA) }
\startdata
GRB000926&2026.136& 280&$ 2.40 \pm 0.14$&$> 2.20$&\\
GRB010222&2056.254& 100&$ 1.51 \pm 0.13$&$> 2.55$&\\
GRB011211&&&$ 1.35 \pm 0.19$&$ 0.70 \pm 0.13$&\\
GRB020813&2260.780& 210&$ 1.71 \pm 0.10$&$ 1.53 \pm 0.05$&\\
GRB030226&2374.461& 340$^a$&$ 0.97 \pm 0.02$&$ 0.32 \pm 0.02$&\\
GRB030323&&&$ 0.75 \pm 0.03$&$ 1.50 \pm 0.03$&\\
GRB050401&&&$ 2.31 \pm 0.26$&$ 1.28 \pm 0.26$&\\
GRB050505&&&$ 1.81 \pm 0.10$&$> 5.92$&\\
GRB050730&1741.553&  25&$ 0.37 \pm 0.01$&$ 0.81 \pm 0.01$&\\
GRB050820&2026.136&  55&$ 1.65 \pm 0.01$&$ 1.51 \pm 0.01$&\\
GRB050922C&1608.451&  89&$ 0.52 \pm 0.01$&$ 0.75 \pm 0.01$&\\
GRB051111&2249.877&  31&$ 0.79 \pm 0.20$&$ 1.35 \pm 0.20$&\\
GRB060418&2576.877&  57&$ 0.66 \pm 0.02$&$ 0.80 \pm 0.02$&\\
\enddata
\tablenotetext{a}{Estimated from Figure~2 of \cite{sbp+06}.  Because the line-profiles are partially saturated, 
this value may be considered an upper limit.}
\tablecomments{All EW values reported are rest-frame equivalent widths.}
 
\end{deluxetable}

We also introduce 
a complimentary diagnostic of gas kinematics: 
the rest equivalent width $W=W_{obs}/(1+z)$.  This quantity has kinematic
significance for optically thick (i.e.\ saturated) transitions. 
In these cases, 
the equivalent width is most sensitive to the differential
motions of individual clouds as opposed to their column densities.  
In contrast to \delv, the $W$ value 
represents the velocity field of greater than $90\%$ of the gas  
and can be dominated by clouds with relatively low column density
\citep[e.g.][]{bmp+07}. 
We emphasize that the $W$ value
should only be sensitive to the metallicity and/or \nhi\ value
of the sightline at very low values, i.e.\ when the 
transition is not saturated.  
Below, we demonstrate
that \delv\ and $W$ are correlated but with large scatter.

\begin{figure}
\begin{center}
\includegraphics[height=3.5in,angle=90]{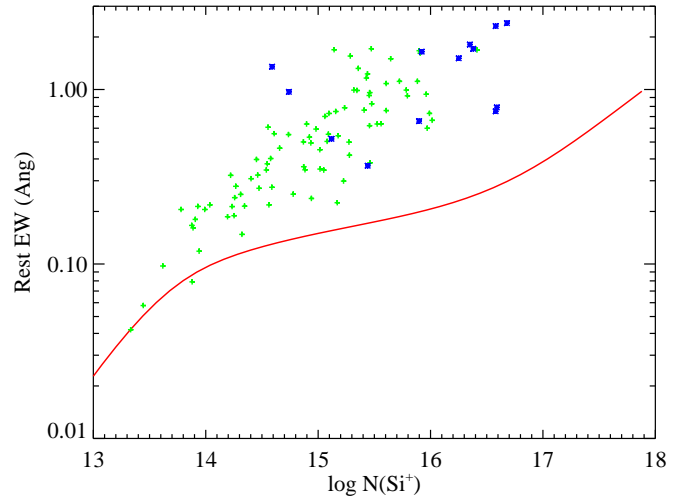}
\end{center}
\caption{Curve-of-growth for the \ion{Si}{2}~1526 transition
assuming a Doppler parameter $b=7\mkms$.  Overplotted
on the figure are the observed total Si$^+$ column densities
and equivalent widths from the QSO-DLAs (green, plus-signs)
and GRB-DLAs (blue stars).  Aside from the few QSO-DLAs
with very low \wsi, the equivalent widths are dominated
by the contribution from multiple clouds with a broad
range of column densities.
}
\label{fig:cogsi}
\end{figure}

We have measured rest equivalent widths from 
the \ion{Si}{2}~$\lambda 1526$ transition (\wsi)
for low-ion gas and the \ion{C}{4}~1548
transition for high-ion gas.  The \ion{Si}{2}~$\lambda 1526$ transition
saturates at $\N{Si^+} \approx 10^{14} \cm{-2}$
which is 10$\times$ lower than the Si$^+$ column 
density for any GRB-DLA sightline and $2\times$ lower than
nearly every QSO-DLA.  
Therefore, this measure should
be relatively insensitive to the gas metallicity or \nhi\ value.
This point is emphasized in Figure~\ref{fig:cogsi} where we 
show the curve-of-growth for the \ion{Si}{2}~1526 transition
and the observed total Si$^+$ column density and equivalent
widths for the QSO-DLAs and GRB-DLAs.
It is evident that the equivalent width has significant
contributions from multiple, lower column density clouds.
Because \ion{C}{4} is a doublet, the \ion{C}{4}~1548 transition will
blend with \ion{C}{4}~1550 when 
$W_{\rm 1548} \gtrsim 2$\AA.  In these few cases, we adopt
a lower limit for \wciv.
The \wsi\ and \wciv\ values of the GRB-DLA are summarized in
Table~\ref{tab:kin}.  We also list the metallicity [M/H]
measurement (or lower limit) for each GRB-DLA 
\citep[see][for details]{pcd+07}.

\section{RESULTS}
\label{sec:results}

In this section, we present the observational results. 
The next section will discuss the implications.

\begin{figure}
\centerline{\epsfxsize=\hsize{\epsfbox{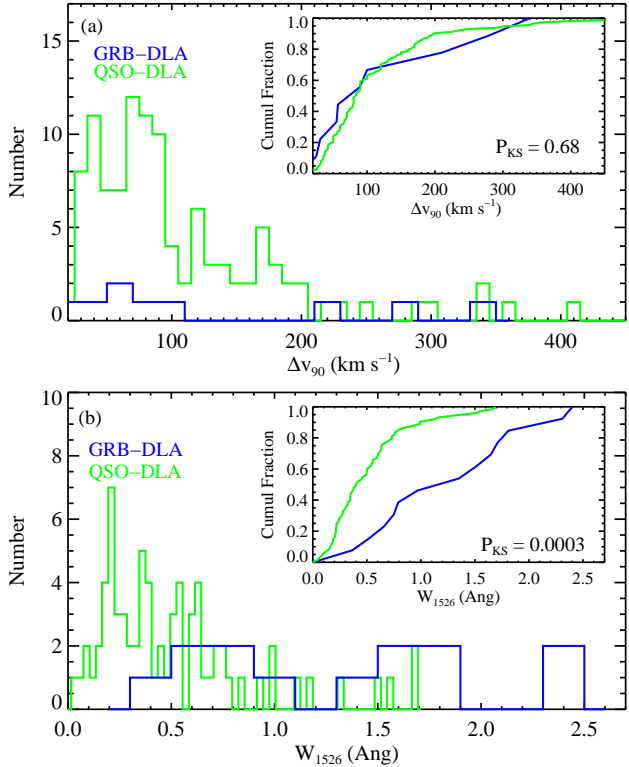}}}
\caption{Histograms of low-ion kinematic statistics for the GRB-DLAs
(dark blue) compared against a statistical sample of QSO-DLAs (light green).
The upper panel shows the results for the \delv\ statistic, the velocity
interval containing 90$\%$ of the optical depth of the low-ion gas.
The two samples have similar distributions with median $\mdelv \approx 80\mkms$
and a significant tail to several hundred \kms.  The lower panel
shows the equivalent widths of the \ion{Si}{2}~1526 transition.
In contrast to the \delv\ statistic, the GRB-DLAs exhibit systematically larger \wsi\ 
values than the QSO-DLAs.  In both panels, the inset shows the cumulative
distributions of the systems, normalized to unity, and the 
$P_{KS}$ value gives the probability that the two distributions
are drawn from the same parent population.
}
\label{fig:hist}
\end{figure}

\subsection{\delv\ and \wsi\ Distributions}

Figure~\ref{fig:hist}a presents the velocity width (\delv) distributions
for the GRB-DLAs and for the statistical subset of QSO-DLAs.
The two distributions are similar;  the majority of galaxies
have $\mdelv < 100 \mkms$ and each sample shows a significant tail out to
several hundred \kms.  A two-sided Kolmogorov-Smirnov 
test rules out the null hypothesis that the two distributions are drawn
from the same parent population at only 32\% c.l.
In the current GRB-DLA sample, there is a higher fraction of 
systems with $\mdelv > 200 \mkms$ than the QSO-DLA distribution.
The difference is not statistically significant at present, but future
observations may reveal a systematic difference in this respect.

Because the GRB are embedded within the interstellar medium 
(Figure~\ref{fig:cartoon}),
the velocity widths are likely an underestimate of the 
value that one would derive by extending the sightline through the
other side of the galaxy.  
If the kinematics are dominated by random motions or rotational
dynamics, then the median increase in \delv\ should be less than a factor
of two.  To be conservative, we have recomputed the 
K-S statistic after doubling each \delv\ value of the GRB-DLAs.
We find results that are still consistent with the null
hypothesis, $P_{\rm KS} = 0.08$. In this regard, 
the sample of GRB-DLA sightlines have velocity fields 
typical of those observed for QSO-DLAs.  
Under the assumption that \delv\ is dominated by the gravitational 
potential \citep[e.g.][]{hsr98},
these results suggest that the parent populations of GRB-DLA and QSO-DLA 
have similar dynamical masses.

From even the first afterglow spectrum \citep{mdk+97}, it was
evident that the gas surrounding GRB exhibit large
equivalent widths from low-ion transitions like \ion{Si}{2}~$\lambda 1526$
and \ion{Mg}{2}~$\lambda 2796$ \citep{sff03}.
As noted in $\S$~\ref{sec:stats}, 
these large equivalent widths do not require large
\nhi\ or metallicity;  the values are often dominated by the velocity
fields of low column density clouds which do not contribute
to the \delv\ statistic.
Figure~\ref{fig:hist}b presents a histogram of \wsi\ values for the
GRB-DLA compared against the values of the
statistical QSO-DLA subset. 
The mean and median of the GRB-DLA distributions are significantly
larger than the QSO-DLAs and a K-S test rules out the null hypothesis
at $99.9\%$c.l.  
Whereas the \delv\ values are in 
rather good agreement between the two samples, 
the \wsi\ values of the GRB-DLAs are systematically larger.
This is an unexpected and puzzling result
which requires unique velocity fields from low column density
gas along GRB sightlines.

\begin{figure}[ht]
\includegraphics[height=3.5in,angle=90]{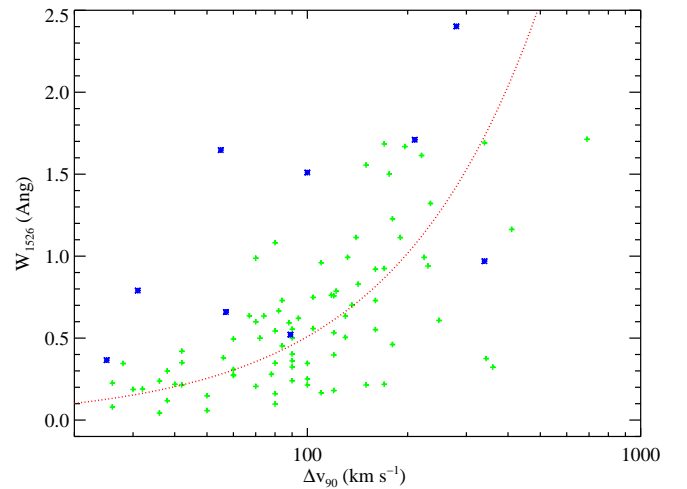}
\caption{Plot of \delv,\wsi\ pairs for the QSO-DLAs (light green)
and GRB-DLAs (dark blue) samples.  The red dashed-line indicates
the \wsi\ value for a boxcar line-profile with velocity width \delv,
i.e.\ $\mwsi = \mdelv * 1526.7066 / c$.
The sightlines that lie significantly above this curve 
must have contributions to \wsi\ from gas at large velocity and
with less than $10\%$ of the optical depth. 
}
\label{fig:dvew}
\end{figure}

\begin{figure*}
\centerline{\epsfxsize=\hsize{\epsfbox{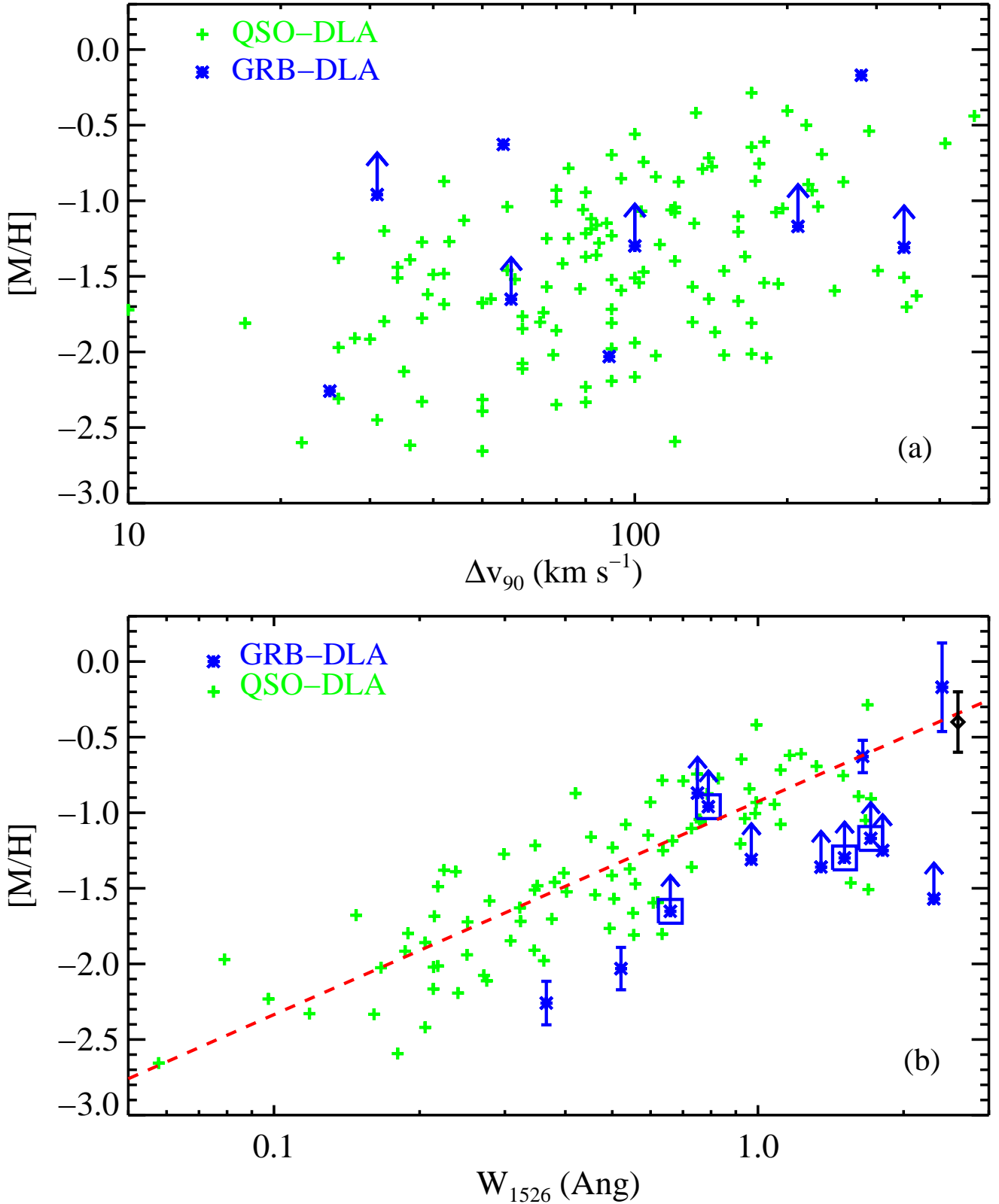}}}
\caption{Plots indicating correlations between the gas 
kinematics and metallicity.  Upper panel shows pairs of \delv, [M/H]
values for the QSO-DLAs (light green plus-signs) 
and GRB-DLAs (dark blue stars).  The GRB-DLA
with lower limits to [M/H] have saturated metal-line
profiles while the points marked with squares do not have
\nhi\ measurements because they have $z_{abs} < 1.6$.  
The QSO-DLAs exhibit a previously
discussed correlation \citep{wp98,ledoux06} and the GRB-DLA
appear to trace the same locus.  The lower panel compares [M/H]
with \wsi.  The QSO-DLA exhibit a remarkably tight trend, described by
the dashed red line: \mwtrend.
The GRB-DLAs also appear to show a correlation which may be offset
or steeper than the QSO-DLA, but this conclusion is complicated
by the preponderance of lower limits.  Finally, the black diamond
shows the [M/H],\wsi\ values for the Lyman break galaxy CB-58 \citep{prs+02}.
Although its correspondence with the QSO-DLA trend may be coincidence,
it is suggestive that large \wsi\ values are related to galactic outflows.
}
\label{fig:mtl}
\end{figure*}

We will argue below that the \delv\ statistic in GRB-DLAs is 
dominated by the ISM of the GRB host galaxy.
It is our expectation that this also holds true for QSO-DLAs.
We will also argue that the \wsi\ values 
result from motions independent of the neutral ISM.
If other environments contribute to \wsi, then
one predicts \delv\ to be only loosely correlated with \wsi.  
Figure~\ref{fig:dvew}, which plots \delv,\wsi\ pairs
for the GRB-DLAs and QSO-DLAs, indicates that this is the case.
The dashed line traces the predicted \wsi\ value for a fully
saturated (i.e.\ boxcar) line-profile that has velocity width
\delv.  Although the QSO-DLAs roughly follow this line,
the GRB-DLAs tend to lie significantly above it.
Regarding the QSO-DLAs, we interpret the results in Figure~\ref{fig:dvew}
as evidence that many sightlines penetrate gas with velocity fields
that are distinct from the majority of gas.
Furthermore, this fraction appears to increase with \delv.
In the GRB-DLAs, nearly every sightline shows a substantial
contribution to \wsi\ from gas that does not dominate the total optical
depth.
In $\S$~\ref{sec:discuss}, we will discuss the origin and nature
of these velocity fields in greater detail.

\subsection{Kinematic Correlations with Other Properties of the ISM}
\label{sec:corr}

We have considered the trends between kinematic characteristics
and other physical properties of the ISM.   
We first examined whether the GRB-DLAs follow the observed 
trends for QSO-DLAs
between gas kinematics and metallicity  \citep{wp98,ledoux06,mcw+07}.
Figure~\ref{fig:mtl}a presents [M/H] values against the \delv\ 
statistic for the
GRB-DLAs and the full QSO-DLA sample.
Note that the GRB-DLA data points with lower
limits to [M/H] have $z_{abs}<1.6$ and an assumed \ion{H}{1} column
density $\mnhi = 10^{22} \cm{-2}$.  Therefore, these [M/H] values
(reported as lower limits because their metal-line profiles are 
saturated) should be considered cautiously.
The QSO-DLAs shown in Figure~\ref{fig:mtl}a exhibit the
correlation reported by previous authors \citep{wp98,ledoux06}.  
It is not a tight trend, however, and we suspect several physical
factors (e.g.\  a mass/metallicity relation; variations with sightline 
impact-parameter and galaxy inclination) 
contribute to produce the observed distribution.
Similar to the \delv\ distribution (Figure~\ref{fig:mtl}a), 
we find that the GRB-DLAs track the locus defined by the QSO-DLAs.  
Adopting the upper limits to [M/H] as values, we report a 
Pearson correlation coefficient of 0.4.
We conclude that there is only tentative evidence for a correlation
between \delv\ and [M/H] for the GRB-DLA.

In Figure~\ref{fig:mtl}b, we present the \wsi, [M/H] pairs
for the two populations.  The QSO-DLA data exhibit a remarkably
tight correlation.  These are are well described by a power-law,
\begin{equation}
{\rm [M/H]} = a + b \log(W/1{\rm \AA}) \;\; ,
\label{eqn:mwsi}
\end{equation}
with best-fit parameters, $a=\aval$ and $b=\bval$.
For this fit, we have assumed equal weights for all of the data
points and we have restricted the sample to $\mwsi < 1.4$\AA\ to 
ignore the `outliers' at large \wsi\ value.  If we include
all of the data points (with equal weighting), 
we derive $a = -1.00 \pm 0.05$ and $b=1.27 \pm 0.10$.
Although the data scatter about this power-law, 
the trend is considerably tighter than the correlation between
\delv\ and [M/H].  In fact, it is the tightest correlation known
between metallicity and any other property of the QSO-DLAs
\citep[see also][]{mcw+07}.
The small scatter is especially impressive given that
$\approx 0.15$\,dex is expected 
from observational uncertainty in the [M/H] measurements (\nhi\ error).
It is a rather surprising trend because the kinematics
of the gas that determines [M/H] are better described by the
\delv\ statistic yet Figure~\ref{fig:mtl}a shows this correlation
has much greater scatter.
Put another way, the line-profiles with 
$\mwsi > 0.3$\AA\ are highly saturated and should not be expected
to reflect the gas metallicity (Figure~\ref{fig:cogsi}).  

This is not to imply that the \wsi\ values are strictly 
independent of [M/H].  For example, an optical depth profile
characterized by shallow, decreasing wings extending 
to large velocity (e.g.\ a Lorentzian profile) 
would yield larger \wsi\ for larger $\N{Si^+}$.
On the other hand, a velocity field with a
sharp cutoff (e.g.\ a Gaussian random field) 
would be insensitive to [M/H].
Furthermore, the peak optical depth value is a function
of both [M/H] and \nhi, and the QSO-DLA data presented in 
Figure~\ref{fig:mtl}b span a factor of $\approx 1.5$\,dex
in \nhi\ value. 
Therefore, the results on the QSO-DLAs in Figure~\ref{fig:mtl}b
indicate an underlying physical mechanism which causally connects
[M/H] and \wsi.
The GRB-DLAs also show a trend between \wsi\ and [M/H], although
its characterization is complicated by the many lower limits to [M/H].    
A qualitative assessment of the
current results suggests either an offset between the trends
for the two populations and/or a steeper power-law for the GRB-DLAs.
We will return to these issues in $\S$~\ref{sec:dismtl}.

Irrespective of the physical origin of the correlation,
the results presented in Figure~\ref{fig:mtl}b indicate one may use
\wsi\ as a proxy for metallicity, especially if applied to a sample of data
points instead of individual systems.   We find that 
a sample of at least 10 QSO-DLA systems gives an average
logarithmic metallicity derived from the observed \wsi\ values and
Equation~\ref{eqn:mwsi} that lies within 20$\%$ of the actual value for
95$\%$ of random trials.
Emboldened by this result, we apply 
equation~\ref{eqn:mwsi} to the GRB-DLAs but offsetting [M/H]
by $\delta_{MH}$ to allow for an offset between the GRB-DLA 
and QSO-DLA trends.
We derive an average logarithmic metallicity for the GRB 
of $<{\rm [M/H]}> = -0.8 + \delta_{MH}$.

\begin{figure}
\includegraphics[height=3.5in,angle=90]{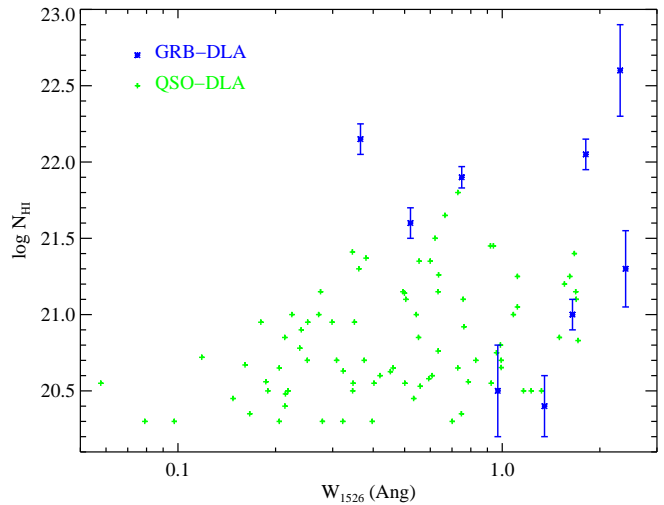}
\caption{Pairs of \wsi, \nhi\ values for the QSO-DLAs (light green)
and GRB-DLAs (dark blue).   Regarding the QSO-DLA, systems with larger
\wsi\ show larger \nhi\ value on average, yet the correlation 
shows large scatter.  Similarly, the GRB-DLA sample is 
characterized by a large scatter but without any
obvious trend.
Although the \nhi\ value and metallicity [M/H] contribute equally
to the Si$^+$ column density and presumably \wsi,
the latter is tightly correlated only with metallicity.
}
\label{fig:ewnhi}
\end{figure}

We have also examined trends between \wsi\ and the \ion{H}{1}
column density \citep[for \delv\ vs.\ \nhi, see][]{wp98,wgp05}.
Figure~\ref{fig:ewnhi} presents the \wsi,\nhi\ pairs for
the QSO-DLAs and GRB-DLAs.  The two quantities are correlated
in that the median \nhi\ value increases with \wsi.
Furthermore, there are no sightlines with low \wsi\
and large \nhi\ value.  For sightlines with $\mwsi > 0.3$\AA,
however, we observe a wide range of \nhi\ values for any given
\wsi\ value.  The large observed scatter contrasts with
the [M/H],\wsi\ trends in Figure~\ref{fig:mtl}b, even though
$\N{Si^+}$ is the product of [M/H] and \nhi.
We conclude that only [M/H] is physically tied to the gas
kinematics expressed by the \wsi\ statistic.

\begin{figure}[ht]
\includegraphics[height=3.5in,angle=90]{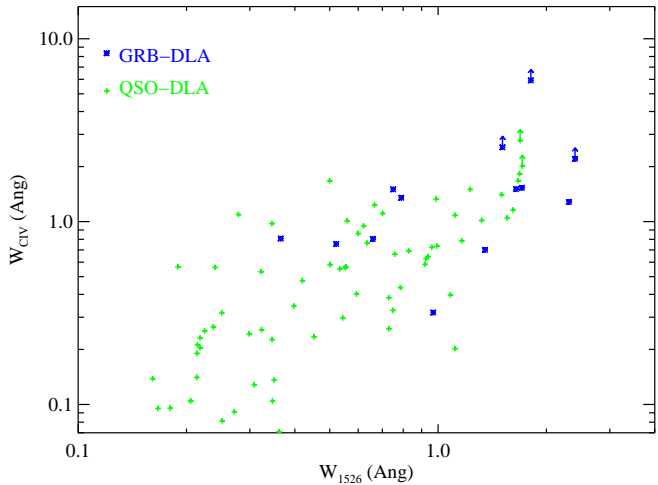}
\caption{Gas kinematics of high-ions (\wciv) compared against
low-ion kinematics (\wsi) for the QSO-DLA (light green) and
GRB-DLA (dark blue) samples.   The two quantities are clearly 
correlated for the QSO-DLAs, whereas the GRB-DLAs exhibit too small
of a dynamic range to robustly exhibit any trend.
}
\label{fig:highlow}
\end{figure}

\subsection{High-ion gas: \wciv}

It is difficult to measure the 90$\%$ velocity interval for \ion{C}{4}
in GRB-DLAs because the \ion{C}{4} doublet is often
saturated.  Therefore, we have
examined the rest equivalent width of the \ion{C}{4}~1548 transition,
\wciv.  These are plotted against \wsi\ in Figure~\ref{fig:highlow}.
Even though unsaturated low-ion profiles do not closely track the
\ion{C}{4} line-profiles in QSO-DLAs 
\citep[i.e.\ component by component;][]{wp00a},
there is a strong correlation between their low and high-ion kinematics.
\cite{wp00b} have interpreted this global correlation
as an indication that the gas traces the same gravitational
potential \citep[see also][]{mps+03}.

The GRB-DLAs also exhibit a significant correlation between
\wsi\ and \wciv.  But, one also finds
that the \wciv\ values of the GRB-DLA are systematically larger
than those for the QSO-DLAs.
Similarly, \wciv\ correlates with the ISM metallicity for both samples.
The correlation between \wciv\ and \wsi\ in the GRB-DLAs
suggests that their velocity fields are at least coupled. 
This assertion is further supported by
the coincidence of \ion{C}{4} and \ion{Si}{2}~$\lambda 1526$ absorption
at some velocities \citep{pcb+07}.  Therefore, investigations on the
velocity field of one ion may give insight to the other.

\section{DISCUSSION}
\label{sec:discuss}

\subsection{Summary of Results}

The results presented in the previous section can be summarized
as follows:

\begin{enumerate}

\item  The distributions of \delv\ values (the velocity interval encompassing
90$\%$ of the gas) from QSO-DLAs and GRB-DLAs are very similar.
Both samples exhibit median $\mdelv \approx 80 \mkms$
with a significant tail to several hundred \kms\ (Figure~\ref{fig:hist}a).

\item  In contrast to the \delv\ statistic, the GRB-DLA exhibit
systematically larger \wsi\ measurements than the QSO-DLAs
(Figure~\ref{fig:hist}b).

\item  Both the GRB-DLAs and QSO-DLAs show a correlation between
\delv\ and metallicity [M/H] but with large scatter (Figure~\ref{fig:mtl}a).

\item  There is a remarkably tight correlation between \wsi\
and metallicity [M/H] for the QSO-DLA sample.  The correlation
is well described by a power-law,
[M/H]~=~$a + b \log(W_{1526}/1{\rm\AA})$ with $a = \aval$ and $b = \bval$.
There is also a correlation observed between metallicity and \wsi\
for the GRB-DLAs. This trend may be offset and/or 
steeper than the QSO-DLA distribution (Figure~\ref{fig:mtl}b).

\item The high-ion and low-ion kinematics traced by \wsi\ and
\wciv\ are correlated (Figure~\ref{fig:highlow}).

\end{enumerate}
We will now explore the origin and nature of the velocity fields
in QSO-DLAs and GRB-DLAs and the
implications for the galaxies which produce them.

\begin{figure*}
\centerline{\epsfxsize=\hsize{\epsfbox{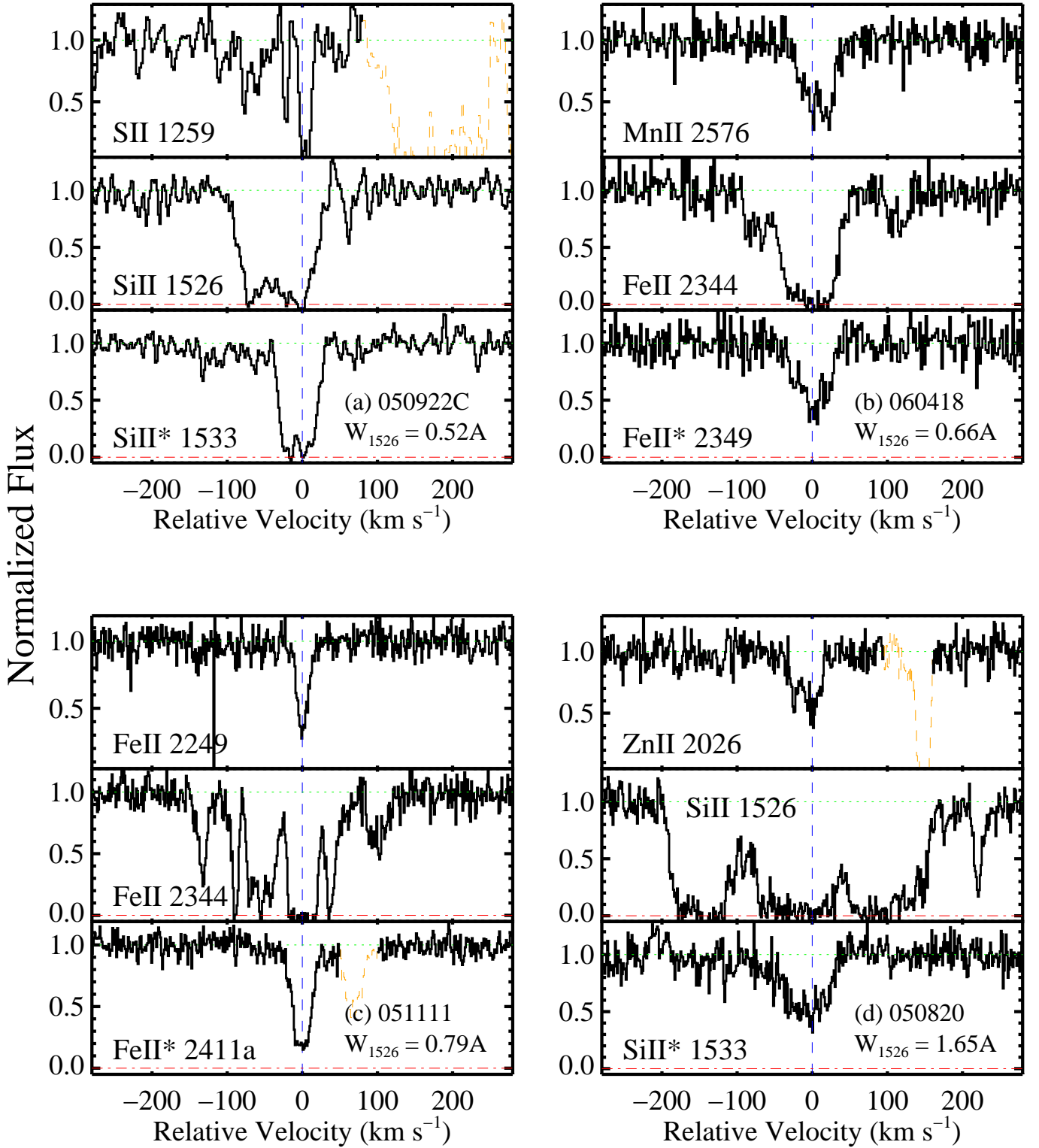}}}
\caption{These panels show one weak line (top), one strong line
(middle) and one fine-structure transition (bottom) for four GRB-DLAs
observed at high spectral resolution: (a) GRB-DLA/050922C;
(b) GRB-DLA/060418; (c) GRB-DLA/051111; and (d) GRB-DLA/050820.
In all four GRB-DLAs, the fine-structure optical depth profile
closely matches that of the weak transition.  Because this gas
must lie within $\approx 1$\,kpc of the GRB afterglow, these
observations indicate the \delv\ statistic is determined by
the velocity field of the ISM surrounding the GRB.
In contrast, there are clouds in the wings of the strong transitions
which do not show significant fine-structure absorption and
must therefore lie at distance $\gtrsim 1$\,kpc from the GRB afterglow.
Because the fine-structure absorption defines the systemic velocity
of the ISM, the unique geometry of the GRB-DLA sightlines
allows one to investigate gas with velocities indicate
of outflow (negative) or infall (positive).
This experiment offers an unprecedented view into the
nature of gas kinematics in high $z$ galaxies.
}
\label{fig:si2}
\end{figure*}

\subsection{The Nature of Galactic Velocity Fields at High $z$}
\label{sec:nature}

In a star-forming galaxy, there are four principal areas
that have unique velocity field characteristics and can contribute
to the \delv\ and \wsi\ statistics:  
(i) the neutral ISM, whose motions are dominated by 
galactic rotation and mild turbulent velocities;
(ii) \ion{H}{2} regions, with velocity fields driven by 
stellar processes (e.g.\ winds, supernovae);
(iii) halo gas clouds, whose kinematics include contributions
from gravitational accretion and virialized motions; and
(iv) galactic-scale outflows driven by supernovae activity in
the ISM \citep[e.g.][]{maclow99} 
and/or heating by galaxy-galaxy mergers \citep[e.g.][]{cjp+06}.
In the following, we will examine our observational results
within the framework of a model where the \delv\ statistic
is dominated by gas in the ambient, neutral ISM 
while halo dynamics determine the 
the \wsi\ and \wciv\ measurements. 
In addition, we will propose that the \wsi\ and \wciv\ statistics
occasionally trace velocity fields representative of
galactic-scale outflows.
%\footnote{
%The term halo gas need not imply 
%absorption at tens of kpc from the ISM.  Indeed, if star
%formation in the ISM is the process which drives the halo
%gas dynamics, then the gas may lie within a few kpc of the disk.
%This gas may be instead referred to as `hot disk' gas \citep{wang}
%or perhaps the warm ionized medium \citep{reynolds,hs00}}.

This two component scenario (ISM+halo gas)
is neither an original nor provocative
interpretation of absorption-line systems arising in galaxies 
\citep[e.g.][]{mm96,kff96,hsr98,mcd99,mps+01,mps+03}.
Indeed, this is also the favored model for gas in the Milky Way 
\citep[e.g.][]{spitzer90,ssl97} and Magellanic Clouds \citep[e.g.][]{lh07}.
And, the model has been invoked for a wide range 
of quasar absorption-line
observations including the impact
parameter distribution of \ion{Mg}{2} systems from their
host galaxies \citep{bb91,lb90,lb92,steidel93} and
\ion{C}{4} absorption by galaxies \citep{clw01,wp00b,mps+03}.
We will demonstrate that our observations support this
model and we will derive new insights into the
gas dynamics of high $z$ galaxies.

Within this disk/halo scenario, the \delv\ statistic traces
the kinematics of the ambient ISM, i.e.,  the optical depth
of the gas external to the neutral ISM contributes 
less than 10$\%$ of the total.  In the context
of QSO-DLAs, this is reasonable because 
(i) the large \nhi\ values that define a DLA may require the
sightlines intersect the ambient ISM;  
and
(ii) one expects halo gas to be predominantly ionized \citep[e.g.][]{mm96}
whereas the majority of gas in QSO-DLA is neutral \citep{vladilo01,pho+02}.
We will now demonstrate that observations of the GRB-DLAs 
directly connect the neutral ISM with the \delv\ statistic. 

Figure~\ref{fig:si2} plots pairs of strong resonance-line transitions
(one weak, one strong)
against a fine-structure transition 
for four GRB-DLAs observed at high spectral resolution.
In all four cases, the fine-structure absorption lines trace
only the gas which dominates the total optical depth, i.e.\ the gas
revealed by the weak transitions.
\cite{pcb06} showed that fine-structure absorption in the
GRB-DLAs is due to UV pumping by radiation from the GRB 
event and its afterglow.  For the GRB-DLAs shown here, the presence of
fine-structure absorption dictates that
this gas is located within $\approx 1$\,kpc of the burst.  Furthermore,
the GRB-DLAs also show 
\ion{Mg}{1} absorption coincident with the fine-structure lines
\citep{pcb+07}. 
This requires the gas to lie at
$\gtrsim 100$\,pc to avoid being photoionized by the GRB
afterglow \citep{pcb06,pcd+07}.
The overall implication is that gas comprising 90$\%$ of the
total optical depth lies between $\approx 100$\,pc and $\approx 1$\,kpc
from the GRB, i.e.\ within the nearby ISM\footnote{The absence of 
H$_2$ absorption in this gas \citep{tpd+07} also indicates we are probing
the ambient ISM and not the local star forming region (i.e.\ molecular cloud).}.   
In summation,
the GRB-DLA observations give empirical confirmation
that the \delv\ values reflect the kinematics of the neutral ISM.

Having established a direct association between the \delv\ values
and the ISM, it is reasonable to assume that these values trace
the rotational dynamics and turbulent fields within high
$z$ galaxies.  This is especially the case
for systems with $\mdelv < 200 \mkms$;
larger \delv\ values may be better interpreted as sightlines
through multiple galaxies 
or even supernovae winds \citep{nbf98}.
Because the turbulent velocities in neutral gas are generally small 
($\sigma_{turb} \approx 10 \mkms$),
sightlines with $\mdelv > 20 \mkms$ should be dominated by
differential rotational along the sightline
or peculiar motions of multiple massive clumps \citep{hsr98,mps+01}.
It is natural, therefore, to compare the \delv\ distributions of
GRB-DLAs and QSO-DLAs in terms of dynamical masses.
Before drawing such conclusions using the \delv\ distributions, however, 
one must examine the effects of differences in sightline 
configuration (Figure~\ref{fig:cartoon}).

The GRB-DLA sightlines are restricted to travel through
approximately half of the galaxy and
are expected to have smaller impact parameter than the 
average QSO-DLA sightline.
The latter point is supported by the higher \ion{H}{1} column
densities, metallicities, and depletion levels observed for
GRB-DLA sightlines \citep{pcd+07}.
These two effects will compete against each other if 
differential rotation explains the \delv\ statistic.
On the one hand, a reduced
sightline leads to an underestimate of 
\delv\ because one does not probe all of the gas along the
sightline.  This effect should be small, however, because even a full sightline
will only probe approximately one quadrant of a rotating
disk \citep{pw97}.  The correction would be larger for a random velocity
field, but still less than a factor of two.
It is only if the velocity fields are dominated by a symmetric
infall/outflow that the effect would be large.  Such velocity fields,
however, are unlikely to describe the neutral ISM of high
$z$ galaxies.  The other geometric effect -- smaller average impact parameter --
will imply larger \delv\ values \citep[e.g.][]{pw97}.  This effect
is likely to dominate and, if anything,  
we expect the GRB-DLAs would be biased to higher \delv\
values than the QSO-DLAs for sightlines penetrating the same parent
population of galaxies.

We conclude, therefore, that 
the observed \delv\ distributions indicate
the galaxies hosting QSO-DLAs and GRB-DLAs have similar mass
distributions.  
There is at least circumstantial evidence in support
of this conclusion. 
In particular,  the median luminosity of
GRB-DLAs host galaxies is low, approximately $L_*/5$ at $z \sim 1$
\citep{ldm+03}.  Although very few QSO-DLA have been successfully
imaged at $z>2$ \citep{mwf+02}, theoretical expectation is that the majority
will also be sub-luminous \citep{hsr00,mps+00,nwh+05} and the
average luminosity at $z<1$ is $L \sim L_*/2$ \citep{cl03}.
Similarly, neither sample has shown evidence for a large fraction
of bright galaxies \citep{ldm+03,cm02}.
Altogether, we conclude that 
the host galaxies of GRB-DLAs and QSO-DLAs have comparable masses.
Adopting recent results on the latter population, 
we infer dark matter halo masses of 
$\lesssim 10^{12} \msol$ for the GRB-DLA \citep{bouche+05,cwg+06,nwh+05}. 

The assertion that GRB-DLA and
QSO-DLA host galaxies have similar mass is immediately challenged
by our observation that GRB-DLA show systematically larger \wsi\ values
(Figure~\ref{fig:hist}b).
To address this issue, we must identify the origin of the
velocity fields contributing to \wsi.  
There are two factors which contribute to the \wsi\ statistic:
the number of clouds intersected by the sightline
and the relative velocities of these clouds.  
The total velocity field is a convolution of the
relative velocity field within each environment giving rise 
to Si$^+$ and the relative
velocities between the environments.  
We emphasize that even highly ionized gas can have trace 
quantities of Si$^+$ that would be revealed by the strong
\ion{Si}{2}~$\lambda 1526$ transition \citep[e.g.][]{pro99}.  
%Therefore, the ambient ISM, the ionized/neutral gas in the galactic
%halo, the clumps entrained in galactic winds,
%and even the ionized gas from the \ion{H}{2} region
%encompassing the GRB will contribute to the \wsi\ statistic.

Returning to Figure~\ref{fig:si2},  we can address the
contribution of various velocity fields to \wsi.
At small \wsi\ values (a,b), the neutral ISM clearly dominates.  
In contrast, 
the two GRB-DLAs with large \wsi\ (c,d) show multiple
clouds with relatively low column density
and no associated fine-structure absorption.
The absence of fine-structure absorption places these clouds
at $\gtrsim 1$\,kpc from the GRB afterglow \citep{pcb06,cpb+07}.  
Note that this is true even though the column densities of these
components are small compared to the total because the resonance
and fine-strucure transitions have similar oscillator strengths.
Therefore, this gas cannot be associated with star-formation
processes local to the GRB (e.g.\ stellar winds, supernovae).
The gas is either within 
the ISM but at much larger distance than the gas that dominates
the total optical depth, or it lies outside 
the ISM altogether (i.e.\ halo gas or the
ISM of a companion galaxy).    We prefer to associate the 
gas with the galactic halo because:
(i) the velocities are large (up to hundreds \kms) 
compared to random and dynamical motions for the neutral ISM of galaxies;
(ii) if the \wsi\ statistic tracked dynamics in the ISM 
then a very large metallicity gradient may be required for 
the \delv\ statistic to be insensitive to these motions;
(iii) there is high-ion absorption (e.g.\ \ion{Si}{4}, \ion{C}{4};
Figure~\ref{fig:exmpl}) coincident with many of the low
column density Si$^+$ components;
and
(iv) the ionic ratios of some components indicate they
are highly ionized \citep{dz06}.
An example of the last point is revealed in Figure~\ref{fig:exmpl}
by the cloud at $v \approx +200 \mkms$.  This cloud has 
$\log [ \N{O^0}/\N{Si^+}] = +0.50 \pm 0.05$ even though O is at least 1\,dex
more abundant than Si in every astrophysical environment known. 
The only viable explanation for observing 
${\rm O^0/Si^+} \ll {\rm (O/Si)_\odot}$ is that the
gas is highly ionized.

Granted that gas beyond the ISM dominates the 
\wsi\ statistic, the next issue is to examine the nature of
its velocity field.  This may include virialized
motions, gravitational accretion \citep{mm96,mcd99}, and outflows from
galactic feedback \citep{dlm03,cjp+06}.  The latter two processes
imply organized flows that would be revealed by the observations
if we knew the systemic velocity of the galaxy. 
For the GRB-DLAs, we can measure the systemic redshift 
from nebular emission
lines \citep[e.g.][]{tgs+07} but the absorption-line data also yield a direct 
measurement: the line centroid of the fine-structure transitions.
Because this gas occurs within $\approx 1$\,kpc for the 
GRB-DLA, it establishes the velocity of the neutral ISM.
Examining the line-profiles of Figure~\ref{fig:si2} (c,d) in this 
light, we note that the Si$^+$ gas shows both positive and
negative absorption in the case of GRB-DLA~050820 but
predominantly negative velocities for GRB-DLA~051111.
Therefore, one example exhibits motions reflective
of a random (i.e.\ virialized) velocity field and the other
inspires an outflow interpretation \citep[see also][]{tgs+07}.
It is too early to draw generic conclusions about the velocity
fields of GRB-DLAs, but future observations akin to Figure~\ref{fig:si2}
will offer an unprecedented view into the nature of gas kinematics
at high $z$.

Let us now address the offset in the \wsi\ distributions between
the GRB-DLAs and QSO-DLAs (Figure~\ref{fig:hist}b).  
Could the offset be explained by differences in sightline configurations?
As with the \delv\ values, there are competing effects.
The bias to smaller impact parameter will only be important
if the Si$^+$ velocity fields have significant
variations on kpc scales.  This could 
occur if the majority of gas is located near star-forming regions,
i.e.\ the origin of the GRB sightline.
If the Si$^+$ gas is distributed 
throughout the halo (i.e.\ tens of kpc), then an impact parameter 
bias should be unimportant.
Meanwhile, if the velocity field has significant contribution from infall 
or outflow then restricting the GRB-DLA sightlines to originate
from within the ISM leads to a systematic underestimate.
Therefore, sightline geometry will bias
the \wsi\ statistic to larger values for the GRB-DLA only if
the velocity field has significant variations on 
kpc scales and is asymmetric.  
Numerical simulations do indicate gas being accreted tends to
fall in along radial orbits (A. Dekel, priv.\ comm.).  This may
imply larger \wsi\ values for sightlines originating at the 
center of the galaxy (GRB-DLAs) as opposed to large impact
parameter (QSO-DLAs).
While differences in sightline configurations may contribute
to the differences in \wsi\ between QSO-DLAs and GRB-DLAs,
we contend that the observed differences in the 
\wsi\ distributions for GRB-DLAs and QSO-DLAs reflect 
actual differences in the gas kinematics of
their host galaxies.

A possible explanation for larger \wsi\ values in GRB-DLA is 
that their host galaxies are significantly more massive.
There are two arguments against this assertion.
First, the \delv\ distributions of the QSO-DLAs and GRB-DLAs 
are similar (Figure~\ref{fig:hist}a).  
While the \wsi\ statistic may be
a better tracer of the gravitational potential (see below), 
the \delv\ statistic should also distinguish between large differences in 
dark matter halo masses.
Second, direct imaging of GRB host galaxies reveal they have
low luminosity \citep{ldm+03,fls+06} and, presumably, low mass.
We consider it unlikely that the QSO-DLA have substantially lower luminosity
(and mass) than that observed for the GRB-DLAs.

The most likely interpretation for the higher \wsi\ values in GRB-DLAs 
is that the kinematics of gas contributing to the \wsi\ statistic
has significant contribution from galactic-scale outflows.
We are guided to this conclusion in part because 
the GRB phenomenon is directly linked to active star formation, i.e.,  
GRB host galaxies have systematically higher current star formation rates
than QSO-DLA galaxies.  
One observes that GRB host galaxies have star formation rates
which are large for their luminosity \citep{chg04}.  This characteristic is
described as a high specific SFR and it separates
GRB host galaxies from the general population of high $z$ galaxies.
Therefore, a scenario where processes related to star formation stir
the velocity fields traced by Si$^+$ gas may naturally account for the
systematic offset in \wsi\ between QSO-DLAs and GRB-DLAs.
These processes may include galactic outflows inspired by major
mergers \citep{cjp+06} or galactic fountains driven by supernovae
feedback \citep{maclow99,dlm03}.
The latter effect may be challenged, however, by the short lifetime
expected for GRB progenitors.
In this respect, the velocity fields of GRB-DLAs may more 
resemble the outflows of the Lyman break galaxies \citep[e.g.][]{prs+02}.
We eagerly await new observations like those presented in 
Figure~\ref{fig:si2} to further test these scenarios.

\subsection{Interpreting the Kinematic-Metallicity Correlations in DLAs}
\label{sec:dismtl}

In Section~\ref{sec:corr}
(Figure~\ref{fig:mtl}), we presented the observed correlations
between the kinematic statistics (\delv, \wsi) and the
ISM metallicity [M/H].  The \delv-[M/H] correlation has been
previously identified for QSO-DLA samples \citep{wp98,ledoux06}.
Although its scatter is substantial, the statistical significance is
high and \cite{ledoux06} report a best-fit power-law of the
form [M/H]$\propto 1.5 \log \mdelv$.  These authors have interpreted
the result in terms of a mass-metallicity relation, i.e.\ the
\delv\ statistic traces the gravitational potential of the
DLA galaxy.  Our results on the GRB-DLA lend further support 
to this interpretation.  In particular, we have shown that the
\delv\ statistic is dominated by material near the GRB, i.e.\ the
neutral ISM.  Although ISM velocity fields are stirred by 
supernovae, 
they are generally dominated by gravitational motions
(e.g.\ rotation, turbulence, accretion). 

A key aspect of the \delv-[M/H] trend is its large
scatter which cannot be attributed to observational
uncertainty.   We interpret the scatter as a 
natural consequence of the QAL experiment.  If the velocity field
is dominated by rotation, then a large scatter results simply from
a range of impact parameters and disk inclinations
\citep[e.g.][]{pw97}.  In this case, only the upper bound of the 
\delv,[M/H] locus would describe the average rotation speed of
galaxies at
a given metallicity.  Similarly, if the velocity field results from
turbulence or accretion, then \delv\ will be sensitive to the number
of clouds intersected by the sightline.  In this scenario, however,
one would
require large variations in the number of `clouds' intersected by
sightlines, which is not entirely supported by the observations
\citep[e.g.][]{pw97}.

In contrast to the \delv-[M/H] trend, the \wsi,[M/H] pairs for
the QSO-DLAs follow an extremely tight correlation (Figure~\ref{fig:mtl}b).
In fact, most of the scatter may be attributed to observational 
uncertainty.  In several respects, this is a stunning result.
The [M/H] values represent the average metallicity along
the entire sightline, which is dominated, of course,
 by the gas corresponding
to the majority of the total optical depth, i.e.,
the gas which defines the \delv\ statistic.  
In contrast, gas which
has negligible contribution to the measured
metallicity can dominate the \wsi\ statistic
(Figure~\ref{fig:si2}).  Furthermore, we have shown (for the GRB-DLAs)
that [M/H] is set by gas local to the ISM of the galaxy whereas
the \wsi\ statistic has significant contribution from gas external
to the ISM.  
Indeed, quasar absorption lines are frequently 
associated with halo gas surrounding a galaxy 
\citep[e.g.][]{bb91,steidel93,clw01,charlton98}. 
Therefore, in the DLAs we reach a rather surprising conclusion
that the local ISM properties (metallicity) are tightly correlated
with the large-scale velocity field.

%[Search for a fundamental plane]

Let us now comment on the origin of the \wsi-[M/H] trend.
The first point to stress is that we derive a similar power-law
as observed for the \delv,[M/H] pairs 
in QSO-DLAs \citep[see also][]{mcw+07}.
We conclude that the
kinematic-metallicity trends have the same physical origin.  
One may gain insight by comparing our trend with the 
velocity-metallicity correlations observed for other
galactic populations.  \cite{dw03} have found the 
velocity dispersion $V$ in dwarf and low-luminosity low surface-brightness
galaxies at $z=0$ correlates with the stellar
mass $M_*$: $V \propto M_*^{0.24}$.  Furthermore, their compilation
shows a tight correlation between 
metallicity and stellar mass, ${\rm M/H} \propto M_*^{0.40}$ implying
a metallicity-velocity correlation: ${\rm M/H} \propto V^{1.66}$.
The correspondence between the kinematic-metallicity trend of
local low-mass galaxies with that of the 
QSO-DLAs is remarkable. 
Although one refers to this trend locally as a mass-metallicity
relation, it is important to stress that simple scaling laws
do not predict the observed trend.  \cite{dw03} have shown
that in a closed-box model the metallicity is independent
of stellar mass; indeed, this follows observations 
for massive, high surface brightness
galaxies \citep[e.g.][]{thk+04}.  
Therefore, \cite{dw03} argued that the mass-metallicity
trend observed for
lower mass galaxies requires supernovae feedback 
\citep[but see also][]{lsc+06,tkg06}.  In this respect, therefore, one
may speculate whether the velocity fields observed in the
QSO-DLAs represent gravitational dynamics or feedback processes.
Let us now consider arguments for and against each interpretation.

There are several arguments supporting the interpretation of 
Figure~\ref{fig:mtl} in terms of a mass-metallicity correlation, i.e.\
that the velocity fields trace the gravitational potential.
First, the observed slopes of the \delv\ and \wsi-[M/H] trends
follow the same trends as local dwarf and low-luminosity low-surface
brightness galaxies \citep{dw03}
whose velocity dispersions presumably reflect
their gravitational potentials.  
Second, the large scatter in the \delv-[M/H] trend can be explained
by variations in impact parameter and inclination through the 
DLA galaxies.  Third, a tight trend is reasonable for
\wsi,[M/H] pairs if the \wsi\ statistic is dominated by halo dynamics.
For virialized motions characterized by a velocity dispersion
$\sigma_{virial}$, \wsi~$\propto n_{cld}^{1/2} \sigma_{virial}$
and the scatter in this quantity will be small for 
a given velocity dispersion $\sigma_{virial}$
if the number of clouds penetrated by the sightline $n_{cld} \gg 1$.
The latter point is supported by the observations 
(e.g.\ Figure~\ref{fig:exmpl}).
Fourth, the \ion{Si}{2}~\,1526 profiles of at least some GRB-DLAs
show both negative and positive velocities relative to the
systemic (Figure~\ref{fig:si2}).  
This is expected for gas that traces a virialized (random) velocity field
(see also Robertson \& Chen, in prep.).
Finally, \cite{mff04} have noted that the only DLAs that have
been successfully imaged at $z \gtrsim 2$ also have high
metallicity, suggesting stellar mass may correlate with enrichment.

There are two counterarguments to interpreting the \delv\ and \wsi\
statistics as tracers of the gravitational potential of individual
dark matter halos.
First, it is may be unrealistic to explain velocity widths that
exceed several hundred \kms\ in terms of a single galactic potential.
At very large \delv\ or \wsi\ values, therefore, we expect
that additional velocity fields contribute.  These could include
galactic-scale outflows but also the peculiar motions of multiple galaxies
along the sightline \citep[e.g.][]{mps+01}.
Second, \cite{bmp+07} have reported an anti-correlation between
\ion{Mg}{2} equivalent width\footnote{In absorbers where one 
observes both the \ion{Si}{2}\,1526 transition and the \ion{Mg}{2}
doublet, one finds the line profiles track each other very closely.
As such, their equivalent widths are highly correlated and roughly
scale as the inverse of their wavelengths: 
$\mwsi \approx W_{2796} \lambda_{1526}/\lambda_{2796}$
(in detail we find $W_{2796} \approx 3 \mwsi$).}
and the cluster-length of these absorbers \citep[a subset include DLAs][]{php+07}
with large, red galaxies (LRGs).  These authors have interpreted
this result in terms of a wind scenario: systems exhibiting
larger equivalent widths occur in less massive, star-bursting galaxies.
We note, however, several caveats to their interpretation:
(i) the \ion{Mg}{2} absorbers have $z<1$ and may 
have very different characteristics (star-formation rates,
metallicity, gas content) than the $z>2$ QSO-DLAs that 
are plotted in Figure~\ref{fig:mtl};
%(ii) LRGs occur in overdense regions of the universe with neighboring
%galaxies that are more likely  old, red and dead.
%These environments may be less amenable to galaxies with
%\ion{Mg}{2} absorption \citep[e.g.][]{co00}.
%A more appropriate test, therefore, may be to cross-correlate QAL absorbers
%with massive, star-forming galaxies 
%\citep[e.g.\ LBGs][]{cwg+06}; and
(ii) \cite{zmn+07} find the integrated light of galaxies
corresponding to \ion{Mg}{2} absorbers is brighter for larger
equivalent width systems suggesting larger masses.

Now consider the evidence in favor of interpreting the \wsi\
statistic (and perhaps \delv) as galactic-scale outflows.  
First, as noted above, the trend of decreasing correlation-length
with increasing equivalent width 
offers at least qualitative evidence for a wind scenario \citep{bmp+07}.
Second, the kinematics of some strong \ion{Mg}{2} absorbers
are consistent with that expected for symmetric outflows
\citep{bcc+01}.
Third, if star formation drives galactic-scale outflows, then
one may expect a correlation between metallicity and \wsi.
It may be difficult to precisely derive the observed trend
([M/H]~$\propto 1.5 \log \mwsi$)
from first principles, but we cannot rule out such a scenario.
Third, we find the Lyman break galaxy \cb\ \citep{prs+02}
lies directly on the \wsi-[M/H] trend of the QSO-DLAs (black diamond in 
Figure~\ref{fig:mtl}b).  
%Before concluding our discussion of Figure~\ref{fig:mtl}b, we note
%that the black diamond at [M/H]=$-0.4$ and $\mwsi = 2.6$\AA\ represents
%the Lyman break galaxy \cb\ \citep{prs+02}.
%Although the LBG observation is inherently different because it
%involves absorption from gas covering sources that span a large
%($\gtrsim 1$kpc) region, 
%the observed correspondence is at the least an amusing coincidence.  
Given that the LBG gas kinematics are dominated by outflows \citep{prs+02},
the correspondence between the LBG data and the QSO-DLA trend
is at least suggestive that outflows may play a role in 
shaping the observed correlation.
The comparison with the QSO-DLAs may
not be appropriate, however, because the LBG sightline is restricted
to originate at the central, star-forming region of the galaxy.  This
implies a special impact parameter and one that presumably probes
only half of the total velocity field.  In these respects, a comparison
with the GRB-DLAs is more appropriate and the
correspondence with the QSO-DLA trend may be simple coincidence.

There are several arguments against interpreting the 
\wsi\ measurements as galactic-scale outflows.
First, the similarity in the \wsi\ and \delv-[M/H] trends would
require one to interpret both
statistics in terms of outflows (or invoke coincidence).  
In turn, one would have to
argue that these winds have gas with large neutral fractions
\citep{pho+02} and significant cross-section to gas 
with $\mnhi > 10^{21}\cm{-2}$ \citep[e.g.][]{schaye01}.
Second, we believe a galactic-scale outflow scenario
would be severely challenged
to reproduce the tight correlation observed for \wsi\ and metallicity.
Assuming outflows are generally collimated,
the \wsi\ statistic should exhibit a wide range of values for a
given wind speed as a function of the sightline configuration
(i.e.\ impact parameter and inclination).
%Unless all high $z$ galaxies are driving significant outflows,
%it may be difficult to invoke a scenrio where winds dominate
%the \wsi\ statistic.
Furthermore, the \wsi\ value is given by the internal 
dynamics of the wind (i.e.\ its turbulence) 
because it is a relative measure independent of the systemic velocity.
It is not obvious that this quantity could be tightly correlated with 
metallicity.
Third, there are at least some GRB-DLAs where the \wsi\ velocity
field is not strictly consistent with outflows (Figure~\ref{fig:si2}).
Fourth, mass-metallicity relations have been identified for numerous
galaxy populations 
at a range of redshifts \citep{dw03,thk+04,kk04,sgl+05,esp+06}.
It would be very remarkable for the QSO-DLAs to exhibit a tight
velocity-metallicity trend that was 
unrelated to a mass-metallicity correlation.
Fifth, we question a scenario which requires galactic-scale outflows in
all galaxies tuned to give
both low and large \wsi\ values with such small scatter.
Sixth, comparisons of \ion{Mg}{2} absorption
line profiles with \ion{H}{2} emission-line kinematics
are inconsistent with a wind scenario and support
dynamical motions \citep{sks+02}.
Finally, while observers are quick to invoke outflows to explain
extreme velocity fields, we question whether outflows can reasonably
explain widths approaching and in excess of 1000\,\kms.

%[Normalize our trend using the local trend]

On balance, we currently favor the interpretation of the velocity
fields in QSO-DLAs as being dominated by gravitational motions,
at least for $\mwsi < 1.5$\AA\ and $\mdelv < 200 \mkms$.
At large \wsi\ and large \delv, it is reasonable to expect that
additional velocity fields contribute.
Indeed, an inspection
of Figure~\ref{fig:mtl}b reveals that the QSO-DLA galaxies with largest
departure from the [M/H]-\wsi\ relation occur at large \wsi\ values.
These few QSO-DLAs have especially large \wsi\ for their observed
metallicity and/or low metallicity for the observed \wsi\ value.
We suspect that the former inference is the correct interpretation
and that these systems have contributions to the \wsi\ statistic
beyond the halo gas dynamics.
These QSO-DLAs are the most viable candidates for sightlines
penetrating galactic-scale outflows. 
This interpretation could also explain the frequency of 
GRB-DLAs that lie off the QSO-DLA trend.  As noted in the
previous section, GRB-DLAs have especially large specific
star formation rates.  Therefore, GRB may flag galaxies
at a time when they are most likely to drive galactic-scale 
outflows.
Given the identification of GRB with massive, short-lived
stars, there may not be sufficient time to drive these winds
with supernovae feedback.  Instead, one may need to
consider merger-driven, galactic outflows.

%[Consider the Steidel MgII result + LBG]

\acknowledgments

We wish to thank A. Dekel, A. Maller, D. Lin,  
E. Ramirez-Ruiz,  P. Bodenheimer,
J. Hennawi, S. Ellison, S. Burles, and R. Bernstein for 
valuable discussions.

%\bibliographystyle{/u/xavier/paper/Bibli/apj}
%\bibliography{/u/xavier/paper/Bibli/journals_apj,/u/xavier/paper/Bibli/grbrefs,/u/xavier/paper/Bibli/qal,/u/xavier/paper/Bibli/ism,/u/xavier/paper/Bibli/instrum,/u/xavier/paper/Bibli/abund,/u/xavier/paper/Bibli/lowzgal,/u/xavier/paper/Bibli/cosm,/u/xavier/paper/Bibli/highzgal,/u/xavier/paper/Bibli/fluor}

%\input{../Tables/tab_grbobs.tex}
%\input{../Tables/tab_grbkinsumm.tex}

\clearpage

\end{document}